# *A New Look at The Neutron and the Lamb Shift*

*by*

*Roger Ellman*

Abstract

The rest mass of the neutron is exactly equal to the rest mass of a proton plus that of an electron plus the mass equivalent of the kinetic energy of those two particles after they have electrostatically accelerated from very far apart toward each other to a separation distance of a proton diameter. That fact is either a remarkable coincidence or evidence that the neutron is a combination of a proton and an electron. The calculation of this sheds new light on the nature and significance of the Lamb Shift.

Roger Ellman, The-Origin Foundation, Inc.
320 Gemma Circle, Santa Rosa, CA 95404, USA
RogerEllman@The-Origin.org
http://www.The-Origin.org

# *A New Look at The Neutron and the Lamb Shift*

*by*

*Roger Ellman*

The neutron gives some evidence of being a combination of an electron and a proton. Unlike the case with atomic nuclei, where the presence of multiple protons and their mutual electrostatic repulsion makes the nucleus tend to fly apart except for the nuclear binding energy, an electron and a proton would tend to bind together in mutual electrostatic attraction. No binding energy / mass deficiency would be needed for an electron - proton combination.

This correlates with the neutron mass, which exceeds the sum of the masses of the hypothesized components, a proton and an electron, by $0.000,839,854$ amu (more than the mass of an electron). The neutron has in this sense a negative mass deficiency or binding energy, a mass excess. One might expect this since the act of combining a proton and an electron should also include at least some of the energy of their mutual attraction.

Because of the negative binding energy one would expect the neutron to be unstable, which it is. While the neutron is quite stable in a stable atomic nucleus, where it is affected by its overall nuclear environment, it readily decays into a proton and an electron when in an unstable nucleus. Furthermore, when free of any nucleus the neutron naturally decays into a proton and an electron with a mean lifetime before decay of about $900$ seconds.

Of course modeling the neutron as a combination of a proton and an electron naturally yields the neutron's electrostatic neutrality. The primary traditional objection to the concept stems from the matter wave wavelength of the electron. In that view the wavelength associated with the electron component of the proton-electron combination would be far too large and in direct contradiction to observed cross-sections and wavelengths.

However, that objection would only apply to a "bunch of grapes" concept of the two particles' combination -- their, so to speak, sitting side by side like two peas in a pod. But if the two particles combine more intimately into a new neutron form their waves might also combine more intimately. Figure 1, below, shows the combination of two oscillations at very different frequencies, the higher symbolically representing the proton and the lower, the electron of the proton – electron pair of which a neutron would be composed.

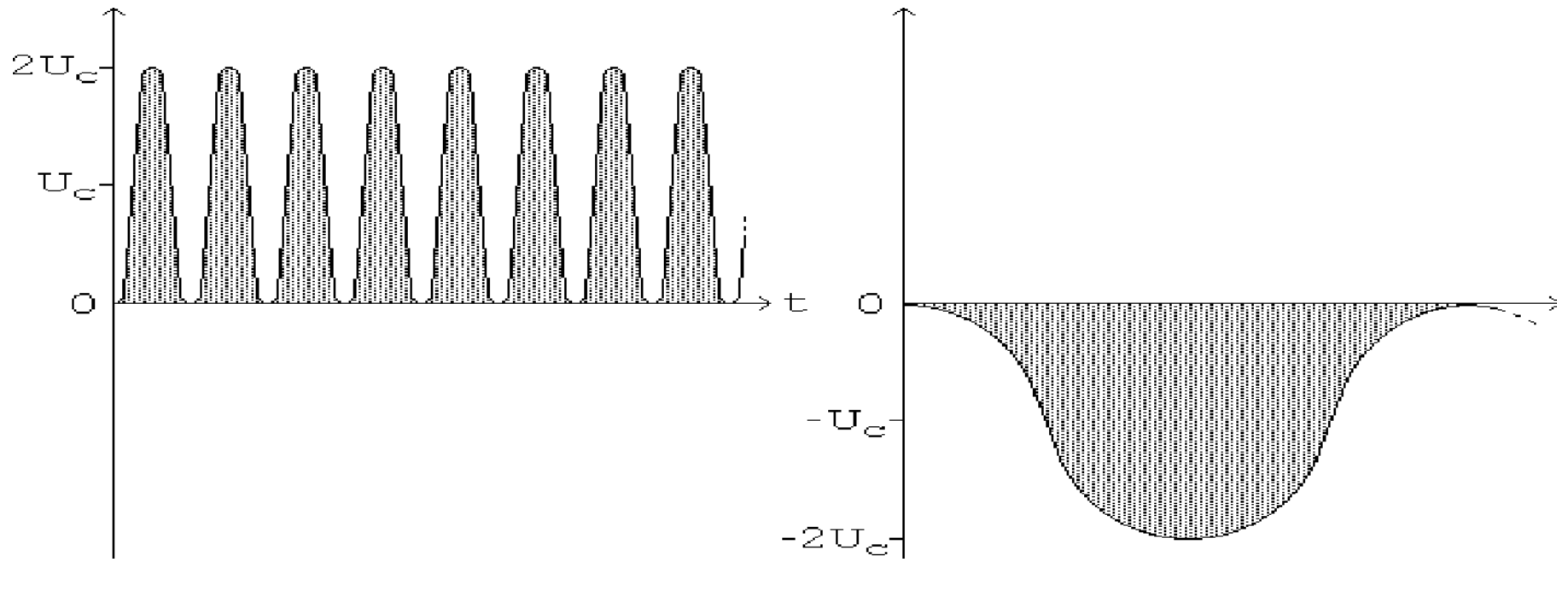

*Figure1(a)*
*Oscillations at Two Different Frequencies*

As in Figure 1(b), below while our eyes can perceive the longer wavelength in the combined wave form (the envelope), the actual oscillation is only at a wavelength essentially that of the shorter input wavelength. The electron's matter wave need not necessarily be a problem.

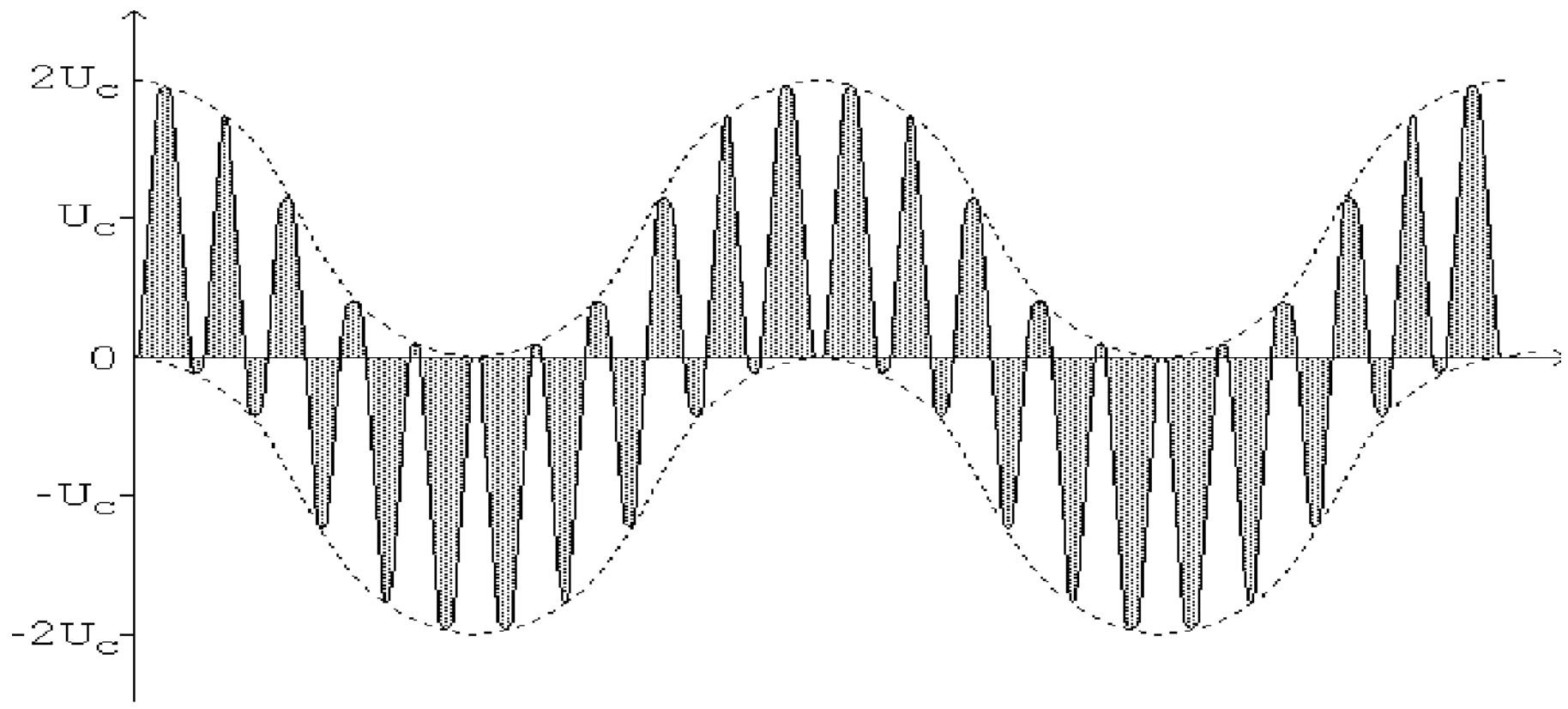

*Figure 1(b)*
*The Sum Oscillation*

The masses of the proton and electron the combination of which is the neutron are not their rest masses even though their combination in the neutron yields the neutron's rest mass. The component masses are the particles' relativistic masses at high velocity. This comes about as follows.

Since a neutron naturally decays into a proton and an electron those decay particles must be emitted at a velocity equal to or greater than their escape velocity. That is, because the proton and electron strongly mutually attract each other electrically, unless they separate at their mutual escape velocities they will immediately re-combine into a neutron.

Put another way, for a neutron to be formed from a proton and an electron the two must come together from the state of being mutually independent of each other. That means that they must mutually accelerate toward each other. In so doing they will each be at escape velocity and have the kinetic energy of that escape velocity at the moment of their combining into the new particle, the neutron.

The portion of the neutron's overall rest mass that corresponds to the component proton and electron's escape velocity kinetic energy is the neutron rest mass less the sum of the proton and the electron rest masses. Using the values for the masses of neutron, proton and electron (per the values recommended in the "1986 Adjustment of the Fundamental Physical Constants" as published by CODATA):

$$(1) \quad \Delta m_n = m_{neutron,\ rest} - [m_{proton,rest} + m_{electron,\ rest}]$$
$$= 1.008,664,904 - \ldots$$
$$\ldots - [1.007,276,470 + 0.000,548,579,903]$$
$$= 0.000,839,854 \text{ amu.}$$

In the "classical" sense escape velocity refers to an object of some mass that is gravitationally bound to some other mass, for example a rocket to be launched from Earth. The force attracting the two objects, the rocket and the Earth, to each other acts on them equally in magnitude and opposite in direction. Consequently, momentums that are equal in magnitude and opposite in direction are imparted to them. Since momentum is the product of mass and velocity, when one object (Earth) is much more massive than the other (the rocket) it may be assumed with negligible error that it (the Earth) is not accelerated and its velocity is negligible. Then all of the kinetic energy is attributable solely to the rocket. That kinetic energy must be equal to the gravitational potential energy binding the rocket to the Earth (the two to each other) for the rocket to escape the Earth's gravitational pull.

However, in the case of a proton and an electron the assumption that only the particle of lesser mass is accelerated and that the other particle's kinetic energy is negligible is not valid. It is not that the electron escapes from the proton; they escape from each other. Or, it is not that the electron falls toward the proton; they fall toward each other. The kinetic energy *(KE)* of each is involved and the sum of the kinetic energies must equal or exceed the binding potential energy *(PE)* for the velocities to be at or in excess of escape velocity.

The analysis (in SI units) is as follows (where $r$ is the closest separation between the escaping objects or particles).

(2) Gravitational | Electrostatic

| Gravitational | Electrostatic |
|---|---|
| Rocket [R] escapes from from Earth [E] | Proton [p] and electron [e] escape from each other |

(a) PE = Force·r

| Gravitational | Electrostatic |
|---|---|
| $PE = \left[G\cdot\frac{m_R\cdot m_E}{r^2}\right]\cdot r$ | $PE = \left[\frac{1}{4\cdot\pi\cdot\varepsilon_0}\cdot\frac{q_p\cdot q_e}{r^2}\right]\cdot r$ |

(b) Final (escape) Kinetic Energy (KE) = Initial Potential energy (PE)

| Gravitational | Electrostatic |
|---|---|
| $KE_R = PE_{total}$ | $KE_p + KE_e = PE_{total}$ |
| $\frac{1}{2}\cdot m_R\cdot v_R^2 = G\cdot\frac{m_R\cdot m_E}{r}$ | No direct solution. A 2nd relationship is: |
| | $\lvert P_p\rvert = \lvert -P_e\rvert$ [P is momentum] |
| $v_{R,esc} = \left[\frac{2\cdot G\cdot m_E}{r}\right]^{\frac{1}{2}}$ | The two relationships above must be simultaneously solved for the velocities. |

For the gravitational case the escape velocity formulation does not involve the mass of the escaping object. In that sense it is independent of the relativistic mass increase with velocity. Furthermore, in the usual cases treating escape velocity of objects (rocketry and astronautics) the velocity never approaches magnitudes at which significant relativistic effects occur.

However, for the electrostatic case the escape velocity formulation must include the masses of the particles, which masses themselves can vary with their velocity. And, in the case of particles, velocities large enough to involve relativistic effects are likely to occur. Therefore, the electrostatic case must be treated relativistically. The simultaneous solution of the two equations, kinetic energy and momentum, is as follows.

*(3)* Momentum

Magnitude of Proton Relativistic Momentum = Magnitude of Electron Relativistic Momentum

$$\frac{m_p}{\left[1-\frac{v_p^2}{c^2}\right]^{\frac{1}{2}}}\cdot v_p = \frac{m_e}{\left[1-\frac{v_e^2}{c^2}\right]^{\frac{1}{2}}}\cdot v_e$$

[$m_p$ and $m_e$ are at rest values.]

Solving the above for $v_p$ the following is obtained.

*(3)* [continued]

$$v_p = \frac{m_e \cdot v_e}{m_p \cdot \left[1 - \frac{v_e^2}{c^2}\right]^{1/2}} \cdot \frac{1}{\left[1 + \frac{m_e^2 \cdot v_e^2}{c^2 \cdot m_p^2 \left[1 - \frac{v_e^2}{c^2}\right]}\right]^{1/2}}$$

*(4)* Energy

Relativistic Energy [As Mass] Is Conserved

$$\left[\frac{KE_p}{c^2} + \frac{KE_e}{c^2}\right]_{gained} = \left[\frac{PE_{total}}{c^2}\right]_{lost}$$

$$[m_{p,v} - m_{p,rest}] + [m_{e,v} - m_{e,rest}] = m_{n,\Delta} \ldots$$

$$\ldots \equiv m_n - [m_{p,rest} + m_{e,rest}]$$

$$\left[\frac{m_p}{\left[1 - \frac{v_p^2}{c^2}\right]^{1/2}} - m_p\right] + \left[\frac{m_e}{\left[1 - \frac{v_e^2}{c^2}\right]^{1/2}} - m_e\right] = m_{n,\Delta}$$

The above expression assumes, it forces, that the excess of the neutron's rest mass above the sum of the mass of a proton plus that of an electron be the KE gained by the two particles in approaching each other from essentially infinite separation distance [per the concept of "escape velocity"]. If, in fact, that is the energy of the two particles at the moment of combining [forming a neutron] then the hypothesis is valid.

The issue here is: how far apart are the proton and electron in their collision paths toward each other when they have the above kinetic masses, $m_{p,v}$ and $m_{e,v}$? For the calculations to be correct, that is for the hypothesis to be correct, their separation distance at that moment must be such that the two colliding particles are exactly at the moment of combining into the neutron. First the velocities, $m_{p,v}$ and $m_{e,v}$, will be calculated by the simultaneous solution of equations *(3)* and *(4)*. Then the separation distance of the two particles at the moment of collision will be determined.

*(5)* Simultaneous Solution

The expression for $v_p$ from equation (3) is substituted for $v_p$ in the denominator of the first term of the expression obtained in equation (4). The resulting expression has only $v_e$ unknown and is solved for that value.

Rather than manipulating that expression a computer aided design program is used to calculate selected trial values of $v_e$ until the desired result for $m_{n,\Delta}$ [$m_{n,\Delta} = m_n - m_{p,rest} - m_{e,rest}$] is obtained.

The results of that process are as follows.

(6) $v_e = 275{,}370{,}263.\ ^m/_s$

$= 0.918{,}536{,}33 \cdot c$

$v_p = 379{,}350.6975\ m/s$

$= 0.001{,}265{,}378 \cdot c$

At those velocities the proton and the electron have total (relativistic) masses of

(7)
$$m_{e,total} = \frac{m_{e,rest}}{\left[1 - \frac{v_e^2}{c^2}\right]^{½}} = 2.529{,}490{,}15 \cdot m_{e,rest}$$

$$= 0.001{,}388{,}308{,}25\ amu$$

$$m_{p,total} = \frac{m_{p,rest}}{\left[1 - \frac{v_p^2}{c^2}\right]^{½}} = 1{,}000{,}000{,}80 \cdot m_{p,rest}$$

$$= 1.007{,}276{,}596\ amu$$

and their sum is the mass of the neutron.

(8)
$$m_{neutron} = m_{p,total} + m_{e,total}$$

$$= 1.007{,}276{,}596 + 0.001{,}388{,}308{,}25$$

$$= 1.008{,}664{,}904\ amu$$

(These calculations assume that the component proton and electron are in a state of zero momentum and zero kinetic energy before being mutually accelerated into each other. It likewise assumes that the resulting neutron has zero kinetic energy and zero momentum because all the components' kinetic energy goes entirely into the neutron's rest mass and the two component's momentums are equal and opposite in direction netting to zero in combination. To the extent that the components do have initial kinetic energy and momentum then either the resulting neutron will have kinetic energy equal to the sum of the components' initial kinetic energies and momentum equal to the net of the two components' initial momenta or some of those quantities may appear in the form of some type of neutrino given off at the time the particles combine.

(Likewise, in describing the decay of a neutron into a proton and an electron, it was assumed that the neutron initially had zero kinetic energy and zero momentum. To the extent that that is not the case then some form of neutrino will account for the kinetic energy and net momentum not accounted for by the decay product proton and electron.)

The remaining issue is: how far apart are the proton and electron in their collision paths toward each other when they have the above kinetic masses ? For the hypothesis to be correct, their separation distance at that moment must be such that the two colliding particles are exactly at the moment of combining into the neutron.

An initial calculation of that separation distance, $r$, is as follows.

*(9)*

$$[\text{Potential Energy}_{\text{As Mass}}] \equiv \frac{PE}{c^2} \text{ and must } = m_{n,\Delta}$$

$$\frac{PE}{c^2} = \frac{q_{proton} \cdot q_{electron}}{4\pi \cdot \varepsilon_0 \cdot r} \cdot \frac{1}{c^2} = [0.000,839,854 \text{ amu}] \cdot [{}^{kg}/_{amu}]$$

$$r = \frac{q_{proton} \cdot q_{electron}}{4\pi \cdot \varepsilon_0 \cdot c^2} = \frac{1}{[0.000,839,854 \text{ amu}] \cdot [{}^{kg}/_{amu}]}$$

The values of all of the quantities except $r$ in the above can be found in the already cited CODATA Bulletin. The result is that the above $r$, the separation distance, is

*(10)* $r = 1.840,636,27 \cdot 10^{-15}$ meters.

Some years ago experiments involving measurement of the scattering of charged particles by atomic nuclei, yielded an empirical formula for the approximate value of the radius of an atomic nucleus to be

*(11)* $\text{Radius} = [1.2 \cdot 10^{-15}] \cdot [\text{Atomic Mass Number}]$ meters

which formula would indicate that the proton radius (atomic mass number $A = 1$) is about $1.2 \cdot 10^{-15}$ meters.

The mass of the proton can be expressed as an equivalent energy, $m \cdot c^2$, and that as an equivalent frequency, $m \cdot c^2/h$, or an equivalent wavelength, $h/m \cdot c$. That wavelength (not a "matter wavelength") for the proton is

*(12)* $\lambda_p = 1.321,408,96 \cdot 10^{-15}$ meters

quite near to the empirical value for the proton radius from equation *(11)*.

Thus the initial calculation of the separation distance of the proton and electron when their kinetic masses are just correct for them to form a neutron, per equations *(9)* and *(10)*, above, results in a separation distance of about 1½ proton radii or equivalent wavelengths, equations *(11)* and *(12)*. That uncorrected result is so close as to essentially validate the hypothesis as it stands.

However, there is more.

The result at equation *(10)* must be corrected for a variation in the magnitude of the classical Coulomb interaction as the charges approach near to each other. The direction of the electrostatic effect of a charge is radial to the charge location. At great distances from a charge all of those radii in a local sample are such a small part of the total spherical Coulomb action that they are effectively parallel. But, near to the charge they all effectively diverge (as, of course, they actually do in all cases). That reduces the electrostatic force and requires the charges to approach each other more closely than to the distance calculated at equation 10 -- in fact to a separation distance of $\lambda_p$ exactly within the limitations of the precision of our data. This develops as follows.

When the two charges are relatively near to each other there is less Coulomb effect because of the radial direction of the Coulomb effect action relative to the charges. Coulomb's law, expressed as potential energy as in equation *(9)*, above, now becomes as follows.

*(13)* $[\text{Potential Energy}_{\text{As Mass}}] = \ldots$

$$\ldots = \frac{[\text{Reduction Factor}] \cdot PE}{c^2} = m_{n,\Delta}$$

*(13 continued)*

$$= [\text{Reduction Factor}] \cdot \frac{q_{proton} \cdot q_{electron}}{4\pi \cdot \varepsilon_0 \cdot r} \cdot \frac{1}{c^2}$$

$$\text{and must} = m_{n,\Delta} = [0.000{,}839{,}854 \text{ amu}] \cdot [^{kg}/_{amu}]$$

But, what is the formulation for the *Reduction Factor*?

For the analysis of the effect of the two charges being so near to each other that the radial divergence of the rays is significant the illustration and dimensions of Figure 2, below, are used. In order to be useful the figure is greatly exaggerated, that is $\alpha$, $\beta$, $d$ and so forth are actually too minute to be seen in an unexaggerated figure.

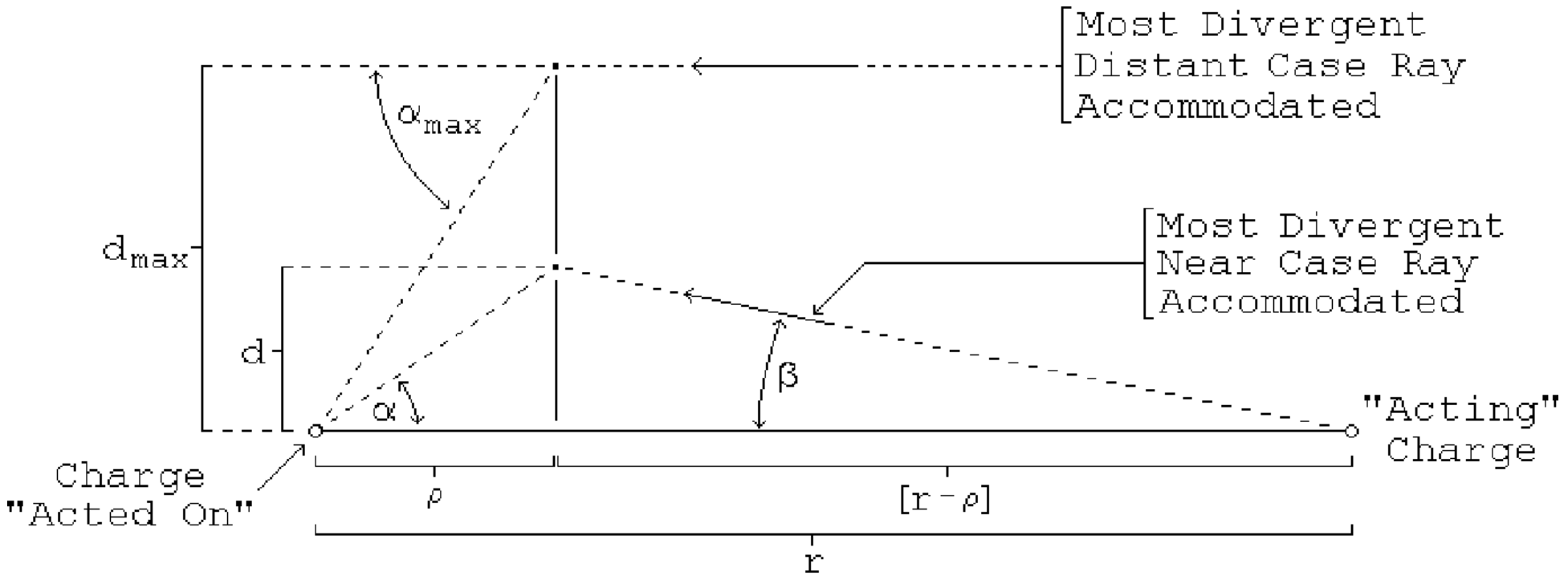

*Figure 2*
*Analysis of Case of Charges Close to Each Other*

Even in the case of charges that are far apart, there is only one single ray that is a straight line from one charge to the other. All other rays of electric field must diverge at least minutely from that one straight ray. Therefore, because of the consistent behavior of the Coulomb Law for distant charges, there is a single constant angle of deviation that accommodates those of the divergent rays that enter into the effect. We need not know for the present purposes what or how that angle is (the entire process is treated definitively in "The Origin and Its Meaning"[1]). But, there must be some such angle which is essentially the same for all cases until the charges are close enough that the radial divergence affects the result. That angle is termed $\alpha_{max}$ in this development.

In terms of Figure 2, for the case of charges near to each other, $\alpha_{max}$ must accommodate both $\beta$ and $\alpha$. When the charges are far apart $\beta$ is essentially zero so that $\alpha_{max} = \alpha$. But, the maximum angle, $\alpha_{max}$, all of which is available to $\alpha$ when the ray source is distant, must, when the ray source is near, account first for removing any ray divergence, $\beta$, with any remaining balance left for $\alpha$.

*(14)* $\alpha + \beta = \alpha_{max}$

(The quantity $\rho$ is needed in order for the concept of $\alpha_{max}$ to have meaning; the angle is pointless without defining where it acts. For charges that are far apart $\alpha_{max}$ and $\rho$ are of no significance. When near effects are operating $\rho$ is at $^r/_2$, half-way between the charges. The concept of $\rho$ is only included here for the initial purpose of presenting in the above Figure 2 the comparison of the near and distant cases.)

The *Reduction Factor* depends upon the reduction of $d$ (of Figure 2) relative to $d_{max}$, that is the ratio $^d/d_{max}$ which quantity is developed as follows.

The angles $\alpha$, $\alpha_{max}$, and $\beta$ are so small that their respective tangents equal their respective angles. Therefore, from the figure

(15)

$$\mathrm{Tan}[\alpha_{max}] = \alpha_{max} = \frac{d_{max}}{\rho}$$

$$\mathrm{Tan}[\beta_{max}] = \beta_{max} = \frac{d_{max}}{r - \rho}$$

$$\mathrm{Tan}[\alpha] = \alpha = \frac{d}{\rho}$$

$$\mathrm{Tan}[\beta] = \beta = \frac{d}{r - \rho}$$

From which

$$\alpha = \frac{d}{d_{max}} \cdot \alpha_{max}$$

$$\beta = \frac{d}{d_{max}} \cdot \beta_{max}$$

Then, substituting the above results into equation *(14)* the following is obtained.

(16)

$$\alpha_{max} = \alpha + \beta = \frac{d}{d_{max}} \cdot \alpha_{max} + \frac{d}{d_{max}} \cdot \beta_{max}$$

From which

$$\frac{d}{d_{max}} = \frac{\alpha_{max}}{\alpha_{max} + \beta_{max}}$$

However, $\alpha_{max}$ is a constant quantity (from the consistent Coulomb behavior when the charges are far apart) as is $d_{max}$.

(17)

$$\alpha_{max} = [\text{A Constant}] \cdot d_{max} \equiv \chi \cdot d_{max}$$

Substituting for $\alpha_{max}$ of equation *(16)* with equation *(17)* and for $\beta_{max}$ of equation *(16)* with $\beta_{max}$ of equation *(15)* the *Reduction Factor* sought for equation *(13)* is obtained. It is the $d/d_{max}$ of equation *(18)*, below.

(18)

$$\frac{d}{d_{max}} = \frac{\chi \cdot d_{max}}{\chi \cdot d_{max} + \frac{d_{max}}{r - \rho}} = \frac{1}{1 + \frac{1}{\chi \cdot [r - \rho]}}$$

$$\text{Reduction Factor} = \frac{1}{1 + \frac{1}{\chi \cdot [r - \rho]}}$$

The form of this effect is depicted graphically in Figure 3, on the following page. This effect is also the cause of the *Lamb Shift*. The Lamb Shift is an extremely fine or slight shifting to higher energy of some of the orbital energy levels of Hydrogen. The Lamb Shift affects orbital electrons that are closer to the atomic nucleus more than those which are more distant; that is, the Lamb Shift is greater as $r$ is smaller. For that reason, it produces a detectable affect principally on the electrons of the inner orbital shells $[n = 1$ or $n = 2]$.

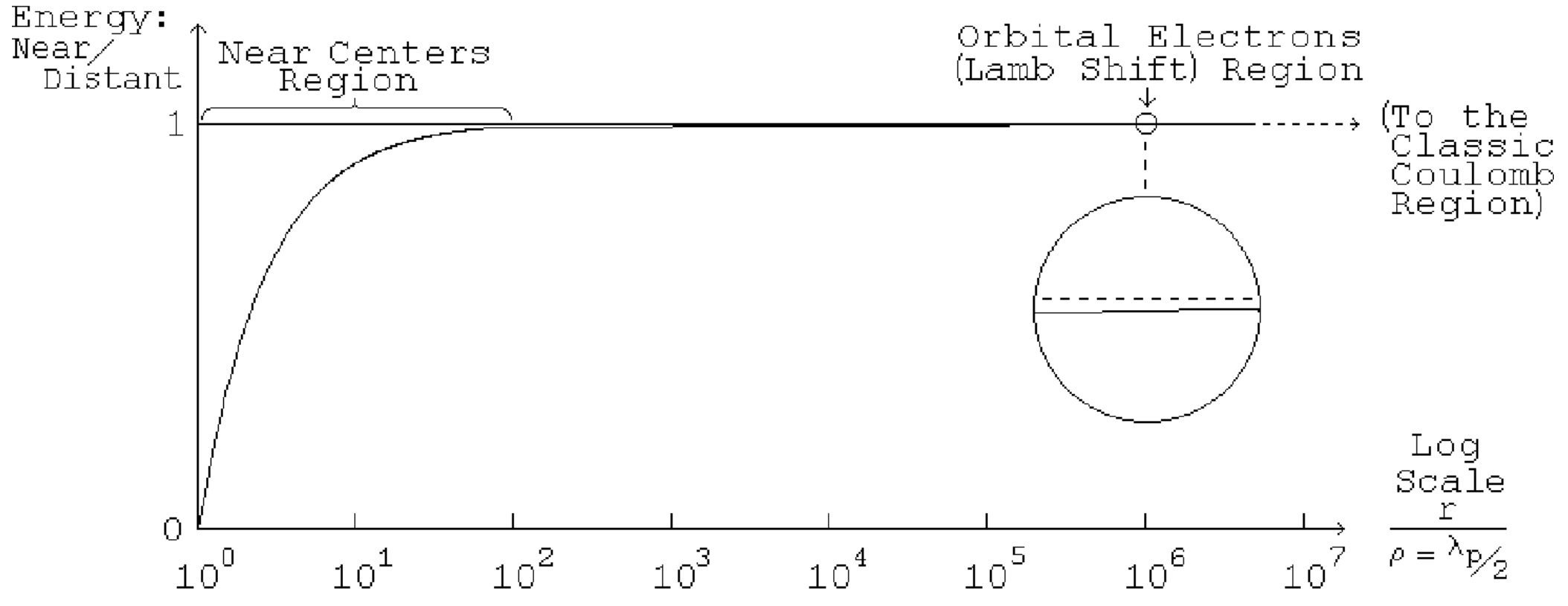

*Figure 3*
*Coulomb Effect* *Reduction Factor* *When Charges Are Near to Each Other*

The Lamb Shift was attributed to "radiative coupling of the electron to the zero point fluctuation of the vacuum". What that means in plain language is as follows. Heisenberg showed that measurement precision is limited because the information extraction process must change the datum while measuring it. 20th Century physics has (questionably) extended that to the attribution of a real uncertainty, not merely one of measurement limitation. Then, the zero of the vacuum would also not be precisely zero but a fluctuation in the Heisenberg uncertainty amount about zero. The Lamb shift was attributed to orbital electron interaction with that fluctuation.

The effect, the Lamb Shift, is actually caused by the reduction in the negative potential energy due to the orbital electron being near enough to the nucleus that the full Coulomb effect, as when the incoming wave is plane, is slightly reduced as developed above. There being at small values of $r$ marginally less Coulomb attraction, the energy pit in which the electron resides is less deep, which means that its energy is somewhat more than would otherwise be the case. The amount of the effect decreases with increasing $r$ because the reduction in the Coulomb effect decreases as $r$ increases.

The Lamb Shift occurs at much larger values of $r$ (electron orbit radii that are on the order of $r = 10^{-10}\ m$) than the quite small value of $r$ at which the neutron forms from the combining proton and electron (on the order of $r = 10^{-15}\ m$). Nevertheless, the Lamb Shift can be used for an approximate calibration of the above *Reduction Factor*.

The Lamb Shift is depicted in Figure 4, below, for the $\ell = 0$ orbit of the $n = 2$ shell.

The original detection of the Lamb Shift was in the $[n = 2]$ Balmer series where the lines are in the visible light range. A similar shift has been measured in the Lyman series $[n = 1]$, which is in the ultra-violet range, at higher frequencies, shorter wavelengths. The Paschen series $(n = 3)$ is in the infra-red range but at radii such that the effect is minute.

The shift is stated in terms of the wave number (reciprocal wavelength) because the Rydberg expression for the spectral lines is in terms of wave numbers. The amount of the *Balmer Â* shift is $0.033\ cm^{-1}$. That occurs at the $n = 2$ level where the overall level itself has the term value the Rydberg constant divided by $n^2$. The fractional shift is then as follows.

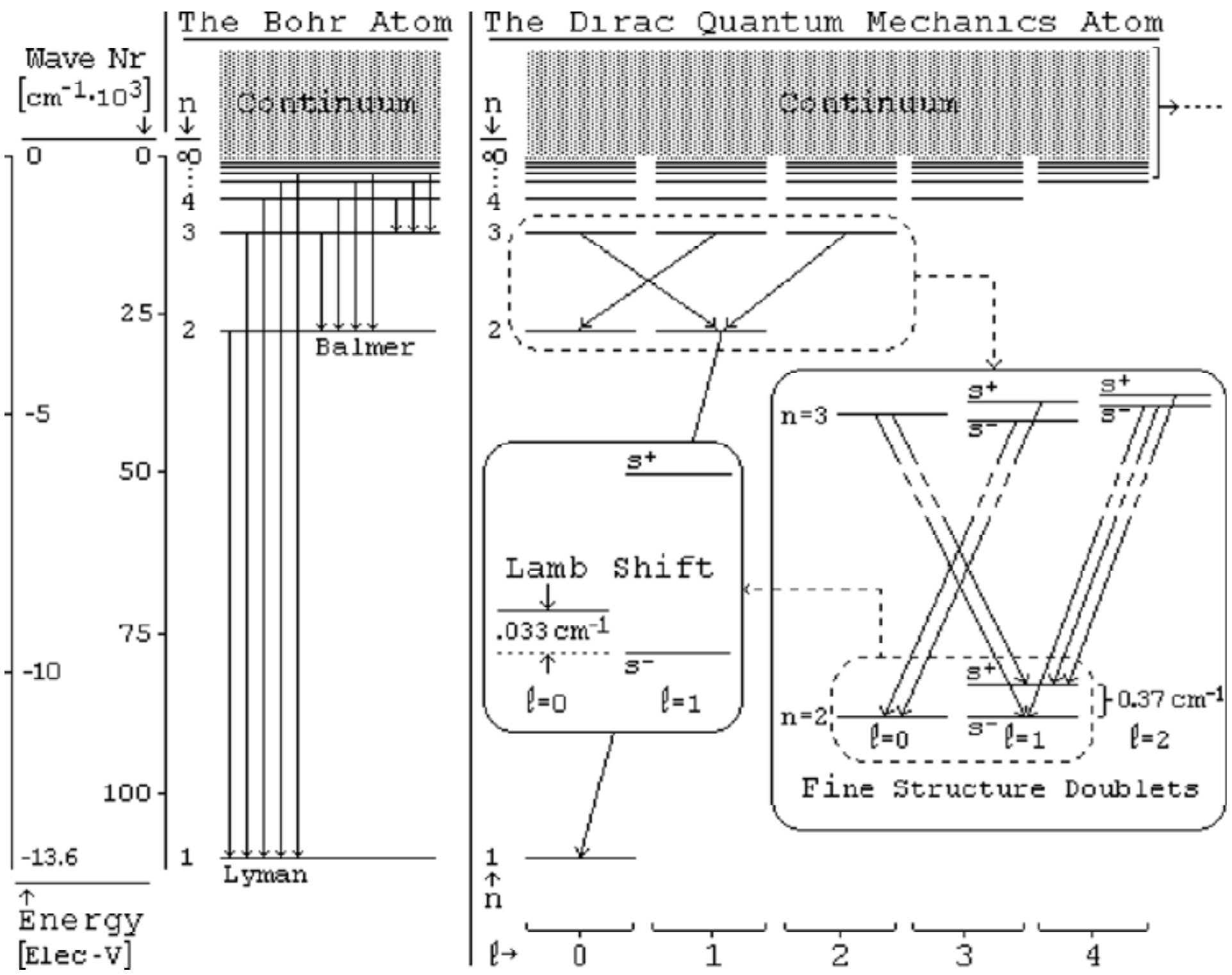

*Figure 4*
*Hydrogen Spectra and the Lamb Shift*

*(19)* $\Delta E = \text{Shift} = 0.033\ cm^{-1}$

$E = \text{Total Wave Number}$

$$= \frac{Ry}{n^2} = \frac{109{,}737.31534}{4} = 27{,}434.3\ cm^{-1}$$

$$\text{Fractional Shift} = \frac{\Delta E}{E} = \frac{0.033}{27{,}434.3}$$

$$= 1.2\cdot 10^{-6}\quad [\text{dimensionless ratio}]$$

The above *Fractional Shift* is the fractional energy change to the "normal" Coulomb potential energy due to the effect of the two charges being near to each other. The *Reduction Factor* as used in this analysis, equation *(13)*, is the net energy after that change, *[1 - the above Fractional Shift]* as follows.

*(20)*

$$\text{Reduction Factor} = [1 - \text{Fractional Shift}]$$
$$= 1 - 1.2\cdot 10^{-6}$$
$$= 0.999{,}998{,}8$$

$$\frac{1}{1 + \dfrac{1}{\chi\cdot[r - \rho]}} = 0.999{,}998{,}8$$

The radius of the $n = 2$ orbit of Hydrogen is $r = 2.1190152 \cdot 10^{-10}$ m. The $\rho$ in the *Reduction Factor* formula is negligible in the case of the Lamb Shift where $r \approx 10^5 \cdot \rho$ and the precision of the Lamb Shift datum is only two significant digits. Equation *(20)* can then be solved for the value of $\chi$ as follows.

(21)
$$\chi = \frac{\text{Reduction Factor}}{r \cdot [1 - \text{Reduction Factor}]} = 3.9 \cdot 10^{15}$$

The general formulation for the *Reduction Factor* is, then, the expression of equation *(18)* with the equation *(21)* value of $\chi$ substituted and $\rho = r/2$. The expression for the potential energy as the proton and the electron approach each other to form a neutron is then equation *(13)* with that *Reduction Factor* substituted. That expression can then be solved for $r$, the $r_{separation}$ with the following result.

(22) $$r_{separation} = 1.3 \cdot 10^{15} \text{ meters}$$

The precision of this result is limited to the two significant digits of the Lamb Shift datum. Nevertheless, it is quite close to the wavelength of the proton oscillation in the neutron per equation *(12)*, $\lambda_p = 1.321,408,96 \cdot 10^{-15}$ *meters*.

Alternatively, if $r_{separation}$ is set at $\lambda_p$ the resulting value for $\chi$ can be calculated and from that the value of $\Delta E$, the Lamb Shift. That calculation gives a Lamb Shift of $.033,611,416\ cm^{-1}$ compared to the actual datum of $.033\ cm^{-1}$.

Two conclusions result from these calculations.

First:

The cause of the Lamb Shift is the change in the magnitude of the Coulomb effect when the two charges are near to each other not the "radiative coupling of the electron to the zero point fluctuation of the vacuum".

Second:

The neutron is the combination of a proton and an electron.